\newcommand{\PaperTitle}{Less is More: Optimizing Probe Selection
Using Shared Latency Anomalies}

\newcommand{\takeaway}[1]{\textcolor{black}{#1}}

\newif\ifshowinfo
\showinfotrue  

\newcommand{\paul}[1]{\ignorespaces}
\newcommand{\fb}[1]{\ignorespaces}

\documentclass[acmsmall]{acmart}

\renewcommand\footnotetextcopyrightpermission[1]{}

\usepackage{graphicx}
\usepackage{subcaption}
\usepackage{multirow}
\usepackage{array}
\usepackage{xcolor}
\usepackage{enumitem}
\usepackage{algorithm}
\usepackage{algpseudocode}
\usepackage{amsmath}
\usepackage{mdframed}
\usepackage{makecell, colortbl}
\usepackage{colortbl}
\usepackage{titlesec}
\usepackage{graphicx}
\usepackage{booktabs}
\usepackage{tabularx}
\usepackage{caption}

\renewcommand{\paragraph}[1]{\noindent\textbf{#1}}

\definecolor{lightgray}{gray}{0.95}

\DeclareMathOperator*{\argmax}{arg\,max}

\titlespacing
\section{0pt}{*1.8}{*1.1}
\titlespacing
\subsection{0pt}{*1.5}{*1.1}
\titlespacing
\subsection{0pt}{*1.3}{*0.8}

\begin{document}

\newmdenv[
  backgroundcolor=lightgray,
  linewidth=0pt,
  leftmargin=10pt,
  rightmargin=10pt,
  innerleftmargin=7.5pt,
  innerrightmargin=7.5pt,
  innertopmargin=5pt,
  innerbottommargin=5pt,
  skipabove=5pt,
  skipbelow=5pt,
]{takeawaybox}

\title{\PaperTitle}

\author{Taveesh Sharma}
\affiliation{%
  \institution{University of Chicago}
  \city{Chicago}
  \country{USA}}
  \email{taveesh@uchicago.edu}  

\author{Andrew Chu}
\affiliation{%
  \institution{University of Chicago}
  \city{Chicago}
  \country{USA}}
  \email{andrewcchu@uchicago.edu}

\author{Paul Schmitt}
\affiliation{%
  \institution{California Polytechnic State University}
  \city{San Luis Obispo}
  \country{USA}}
  \email{prs@calpoly.edu}

\author{Francesco Bronzino}
\affiliation{%
 \institution{\'Ecole Normale Sup\'erieure de Lyon}
 \city{Lyon}
 \country{France}}
 \email{francesco.bronzino@ens-lyon.fr}

\author{Nick Feamster}
\affiliation{%
  \institution{University of Chicago}
  \city{Chicago}
  \country{USA}}
  \email{feamster@uchicago.edu}

\author{Nicole P. Marwell}
\affiliation{%
  \institution{University of Chicago}
  \city{Chicago}
  \country{USA}}
  \email{nmarwell@uchicago.edu}

\begin{abstract}

Latency anomalies—persistent or transient increases in
round-trip time (RTT)—are a common feature of residential Internet
performance. When multiple users simultaneously experience anomalies at the same
destination, it may indicate shared infrastructure issues,
routing behavior, or congestion. However, inferring such shared behavior is
challenging in practice. This is because the magnitude of these anomalies can vary
significantly across devices, even within the same ISP and geographic area,
and detailed network topology information is often unavailable due to platform
limitations or privacy constraints.

In this work, we study whether devices that experience a shared latency anomaly
observe similar changes in RTT magnitude using a topology-agnostic approach.
Using a four-month dataset of high-frequency RTT measurements from 99
residential probes in Chicago, we detect shared anomalies and analyze their
consistency in amplitude and duration without relying on traceroutes or
explicit path information. Building on prior change-point detection
techniques, we find that many shared anomalies affect users similarly in
amplitude, particularly within the same ISP. Leveraging this insight, we
develop a sampling algorithm that reduces redundancy in detected anomalies
by selecting representative devices under user-defined constraints. Our
approach covers 95\% of aggregate anomaly impact with less than half the
total probes used in our deployment. Compared to two baselines, we show that
our approach selects a significantly higher number of unique anomalies at
similar coverage levels. Additionally, our analysis suggests that geographic
diversity can play an important role in selecting probes for a single ISP
even within a single city. These findings highlight the potential of using
anomaly amplitude and duration as topology-independent signals for scalable
monitoring, troubleshooting, and cost-effective sampling designs in
residential Internet performance measurement.

\end{abstract}

\maketitle

\section{Introduction} \label{sec:introduction}

Residential broadband network performance has been an area of active
research and policy interest \cite{sundaresan2011broadband,
paul2023decoding,lee2023analyzing,manda2024efficacy,nabi2024red,
sharma2024beyond}, given its critical role in enabling access to
several aspects of digital life. While aggregate Internet performance
has improved over the years \cite{cisco2020InternetReport,
datanet2023HistorySpeed}, Internet users continue to experience frequent
performance disruptions in the form of low speed, elevated latency,
and packet loss
\cite{manda2024efficacy,gettys2012bufferbloat,clark2022understanding}.
Among these, latency anomalies—transient as well as persistent
increases in round-trip time (RTT)—particularly affect
latency-sensitive applications such as video conferencing
\cite{macmillan2021measuring,sharma2023estimating}, online gaming
\cite{alvarez2023using}, and other real-time services.

Latency anomalies can significantly degrade user experience without
causing complete service outages. Because they do not operate as
binary reachability failures, such anomalies are often invisible to
coarse-grained availability metrics (e.g., loss of connectivity,
prefix withdrawals, or host reachability) commonly used by
traditional outage detection systems \cite{yan2009bgpmon,
renita2017network}. As a result, operators and policymakers may
underestimate both the prevalence and the impact of performance
degradations that affect users without triggering alerts.

Prior research suggests that latency anomalies can arise from routing
changes \cite{carisimo2022jitterbug}, server-side anomalies
\cite{roy2013characterizing}, shared infrastructure bottlenecks
\cite{sharma2023first}, congestion \cite{dhamdhere2018inferring}, or
other network phenomena. Robust techniques have been developed to
detect such events using statistical change-point detection applied
to longitudinal measurements
\cite{luckie2014challenges,dhamdhere2018inferring,carisimo2022jitterbug}.
These studies analyze anomalies at the level of
individual probe-destination pairs or rely on explicit path or
topology information (e.g., traceroutes) to reason about shared
infrastructure. Unfortunately, in many practical residential network measurement
deployments, traceroutes are unavailable due to privacy constraints,
platform limitations, or measurement overhead, leaving operators with
little visibility into shared paths. For instance, Measurement Lab (M-Lab) \cite{mlab2025} collects 
server-to-client traceroutes for speed tests opportunistically, and are not guaranteed to be
available for all measurements. Further, Ookla's Open Data \cite{ookla_open_data_2025} provides 
only aggregated client-side measurements without any path information. Finally, while RIPE Atlas \cite{ripeatlas2025} 
provides access to some traceroute data, it is difficult to find cities with a sufficiently large number of probes to conduct a meaningful within-city analysis. In such settings, it remains
unclear how to identify redundancy among probes or select
representative devices in a principled manner.

\begin{figure}
  \includegraphics[width=0.45\linewidth]{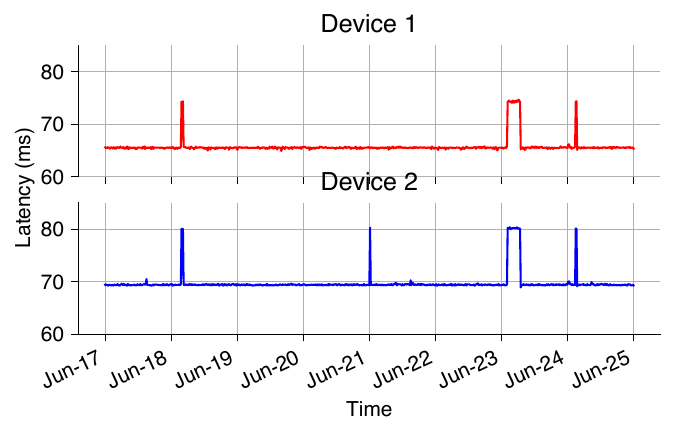}
  \caption{Shared latency anomalies to a Seattle-based public
    measurement server for two different residential AT\&T devices
    located in the same zip code.\paul{minor: generally speaking a
      figure has to be truly incredible to make it to page one. Otherwise
  it's better to move it to page 2,}}
  \label{fig:motivating_example}
\end{figure}

In this paper, we pose a largely unexplored question: \textit{when two
or more residential devices experience a latency anomaly to the same
destination at the same time, do they experience it with similar
magnitude?} Unlike an array of prior work that characterizes network performance at coarse granularities (city- \cite{ozcan2023longitudinal}, metro- \cite{canadi2012revisiting}, AS/ISP- \cite{deng2021comparing}, or country-level \cite{chavula2017insight}), our focus on inter-probe correlation within a city enables a microscopic understanding of how the digital divide manifests in residential broadband networks. Further, we ask this question without assuming access to traceroutes or explicit topology information. Instead, we rely
solely on end-to-end latency measurements and their spatio-temporal
structure to infer shared behavior across probes from a dense urban deployment. This design reflects
the constraints faced by many operational deployments and allows us to
study probe redundancy in a topology-free manner. This question is important for two reasons. First, if latency
anomalies are spatially correlated and exhibit similar amplitude
across devices, then analyzing every probe independently introduces
substantial redundancy. Second, understanding the consistency of
anomaly magnitude across devices can provide valuable diagnostic
signals for localizing faults (e.g., last-mile versus middle-mile
bottlenecks), even in the absence of path-level visibility. If such
consistency exists, measurements from a small number of
representative probes may suffice to characterize the broader impact
of network events. Figure~\ref{fig:motivating_example} illustrates this intuition for a
pair of residential devices located in the same zip code and
subscribed to the same ISP (AT\&T). Both devices experience three
shared latency anomalies to a Seattle-based M-Lab
server during June 2022. The time series are shown after aggregating
minimum RTT measurements into 15-minute bins. The shared anomalies
exhibit similar amplitudes and durations, suggesting that sampling
measurements from either device would capture nearly identical
information about these events and may help localize their root cause.

To explore these questions, we analyze a four-month dataset of
high-frequency RTT measurements collected from 99 fixed residential
probes deployed in home networks across Chicago. Each probe measures
RTT to a common set of destinations at five-minute intervals. We
primarily focus on latency to M-Lab servers
\cite{mlab2025}, whose static IP addresses and global footprint allow
us to detect network-induced anomalies while minimizing noise from DNS
resolution or server-side dynamics. Building on prior work
\cite{carisimo2022jitterbug,dhamdhere2018inferring}, we apply
change-point detection techniques to identify latency anomalies and
analyze the relationship between their temporal overlap and similarity
in amplitude across probes.

We then formulate the problem of probe selection as a maximum weighted
set coverage problem \cite{khuller1999budgeted,nemhauser1978analysis},
where sets correspond to distinct latency anomalies and weights
capture their impact. Unlike prior approaches that rely on inferred
paths or shared links, our formulation operates entirely on anomaly
co-occurrence and impact. It requires no knowledge of the underlying
network topology. To approximate the optimal solution to this NP-hard
problem, we design a simple greedy heuristic that maximizes coverage
of distinct anomalies while minimizing the number of selected devices.

Our results show that shared anomalies often exhibit similar
amplitudes across devices, particularly when users share the same
ISP, and that anomalies with greater temporal overlap tend to have
higher impact. These findings work towards answering a practical question faced by
operators and researchers alike: \textit{can we reduce measurement redundancy
in existing deployments without losing visibility into important
performance degradations?} Reducing redundancy lowers data collection
and processing costs, simplifies downstream analyzes, and mitigates
noise introduced by multiple probes observing the same event. In this
sense, probe selection acts as a form of spatio-temporal down-sampling
that preserves the most operationally relevant signals in large-scale
measurement datasets.

To quantify these trade-offs, we evaluate how many devices are needed
to capture the majority of observed anomaly impact, which we define as the
product of amplitude and duration. We find that capturing 95\% of
total impact requires fewer than half of the probes in our deployment
(44 out of 99). Compared to uniform random selection, which selects 33
probes, our approach detects 2.2$\times$ more unique anomalies toward
a local M-Lab server. We further show that a small amount of
historical data (1--2 weeks) is sufficient to select probes that
continue to provide steady anomaly coverage in the future. Together,
these results indicate that careful, topology-agnostic probe
selection can significantly reduce measurement overhead while
preserving the ability to detect and characterize impactful latency
anomalies in residential broadband networks.

In our knowledge, this is the first work that studies co-occurrence of latency anomalies across a citywide deployment to inform probe selection without relying on path or topology information. We make the following contributions:
\begin{itemize}[leftmargin=*, topsep=0pt, itemsep=0pt]
  \item We show that latency anomalies observed by different
    residential probes often exhibit similar amplitudes and durations
    when they temporally overlap, even without access to path or
    topology information (Section~\ref{sec:shared_elevations}).
  \item We develop a topology-agnostic probe selection algorithm that
    leverages temporal overlap and anomaly impact to identify
    representative devices in longitudinal measurement deployments
    (Section~\ref{sec:algorithm}).
  \item We demonstrate that our algorithm covers 95\% of anomaly
    impact using fewer than half of the probes in our deployment and
    significantly outperforms baseline selection strategies in detecting unique anomalies 
    (Section~\ref{sec:eval-algorithm}).
  \item We show that a small amount of historical data (1--2 weeks) is
    sufficient to select probes that continue to provide steady
    anomaly coverage in the future
    (Section~\ref{sec:relevance}).
\end{itemize}

\section{Background \& Related Work} \label{sec:background}

This section surveys prior work that informs our methodology and
contributions. We begin by reviewing studies on broadband Internet
performance and Internet topology mapping. We then describe
change-point detection techniques for
latency time-series data and detail the foundation of our approach,
including limitations of prior work that motivate our modifications.
Next, we discuss strategies for probe selection in large-scale active
measurement platforms, emphasizing how our data-driven approach
differs from topology-based methods. Finally, we outline the maximum
weighted set coverage problem, a classical optimization problem that
underpins our formulation for selecting representative probes.

\paul{I generally prefer related at the end, but IMC people seem a
  little more traditional than me. Might consider moving to a
  subsection in S2 and calling it ``Background and Related Work'' or
something like that.}
\subsection{Broadband Internet Performance}
Several prior studies have analyzed broadband Internet performance
using large-scale measurement platforms \cite{canadi2012revisiting,
sundaresan2012measuring,chetty2013measuring}. The FCC's Measuring
Broadband America (FCC MBA) program~\cite{fccmba2025} and
M-Lab~\cite{mlab2025} provide insight into ISP-level speed and
latency characteristics for individual connections, including
variation by geography and over time. Sundaresan \textit{et
al.}~\cite{sundaresan2011broadband} show that broadband performance
shows significant deviation from advertised speeds, particularly
during peak hours. More recent research has shifted the focus from
understanding a single end-to-end path to analyzing Internet
performance dynamics for collective user populations. These studies
can broadly be classified into regional performance comparisons
\cite{sharma2022benchmarks,paul2022characterizing}, mapping efforts
\cite{bronzino2021mapping,saxon2022we}, statistical modeling
\cite{lee2023analyzing,jiang2023mobile, sharma2024beyond}, and policy
focused work \cite{manda2024efficacy,
  marques2024challengeanalysisfccbroadband,clark2022understanding,
macmillan2023comparative,mangla2022internet}. While these studies
recognize the importance of understanding the spatio-temporal
dynamics of Internet performance, our work is the first to utilize
shared latency anomalies to reduce redundancy in measurement infrastructure.

\subsection{Internet Topology Mapping}

A number of efforts have been made to better understand Internet topology, both
in regard to its structure (connectivity-based and geographic)
\cite{chengRealNetTopologyGenerator2008,heidemannCensusSurveyVisible2008} and
state
\cite{baltraImprovingCoverageInternet2020,guillotChocolatineOutageDetection2019,zhangPlanetSeerInternetPath2004,katzBassettLIFEGUARDPracticalRepair2012,djatmikoCollaborativeNetworkOutage2013,richterCountingNewPerspectives2016,katz2008studying,richter2018advancing}.
In this space exist a number of high-level parallels to our objectives. One
such objective is improving measurement scalability while retaining fidelity and
resilience against false positives. Hu~\textit{et
al.}~\cite{huGeolocationMillionsIP2012} present an algorithm for geolocating IP
addresses that selects the best few vantage points to measure from while
maintaining accuracy comparable to approaches that rely on more measurements. This effort
is very similar to our work, with vantage points being tantamount to the
residential probe devices described in this paper. Our work differs in that we
attempt to efficiently and accurately measure latency anomalies as compared to
exploring how to scale up IP geolocation algorithms. Various other works also
examine this problem in the context of active
probing~\cite{quanTrinocularUnderstandingInternet2013, quan2014internet,
padmanabhanHowFindCorrelated2019}.


Another parallel objective is determining if similarities exist between
endpoints that share spatio-temporal characteristics, and the implications
of these similarities. Cai~\textit{et
al.}~\cite{caiUnderstandingBlocklevelAddress2010} develop a network address block clustering
technique which reveals that contiguous addresses share qualities such as utilization
and link speed. Similarly, Baltra \textit{et
al.}~\cite{baltraWhatInternetPartial2023} develop two algorithms for detecting
groups of IP addresses with differing connectivity characteristics (islands and
peninsulas), and use these taxonomies to improve the DNSMon tool~\cite{dnsmon}. Our efforts in
this paper hope to answer a similar question, but in regard to residential
Internet performance—do devices that experience the same latency anomaly
experience it at the same magnitude? We detail our findings and their implications
towards answering this question in Section~\ref{sec:eval-algorithm}.

\subsection{Latency Time-series Change-point Detection}
Detecting change-points in latency time-series is a widely used
technique for several applications. These include detecting
congestion and path changes \cite{dhamdhere2018inferring,
luckie2014challenges, carisimo2022jitterbug, hou2021detection},
detecting routing attacks
\cite{deshpande2009online,jubas2021detecting}, adaptive network
management \cite{simon2023change}, and application performance
monitoring \cite{cito2015identifying, fleming2023hunter}. A variety
of statistical and algorithmic techniques have been proposed, ranging
from CUSUM \cite{taylor2000change} and Bayesian change-point models
\cite{xuan2007modeling} to signal decomposition and machine learning
based approaches ~\cite{killick2012optimal, barford2002signal}.

Our methodology for detecting latency anomalies in an RTT time-series
is an extension of Jitterbug \cite{carisimo2022jitterbug}. Jitterbug
first uses Bayesian change-point detection (BCP)
\cite{xuan2007modeling} or Hidden Markov Models (HMMs)
\cite{mouchet2019flexible} to identify a set of candidate timestamps
where a change in latency occurs. Here, a change encompasses both
positive and negative shifts in mean latency. Then, it uses a set of
heuristics to filter positive changes that are less likely to be
caused by a significant network event. The heuristics include
inferring a ``jump" in latency by checking if the difference in means
of consecutive segments is greater than a threshold (the
recommendation is 0.5 ms), and checking whether dispersion in jitter
is greater than a threshold to signal congestion.

While we acknowledge the effectiveness of Jitterbug in distinguishing
congestion events from other network phenomena, we make a number of
modifications to make this approach more suitable for our use case.
First, as the authors note, Jitterbug's range of supported
change-point detection methods do not detect \textit{all} periods of
elevated latency. In our work, we make an attempt to address this
limitation by introducing heuristics and faster detection methods
with greater sensitivity to flag more jumps. Second, we notice in our
experiments that Jitterbug applies change-point detection on the
entire time-series at once, which may result false-negatives. For
example, if change-points are located at the edges of the RTT
time-series, they are likely to be considered as ``normal" behavior
due to limited sample availability around boundaries. Since we are
interested in identifying shared network anomalies across devices, it
is important to ensure that we detect maximum changes in latency.
Finally, Jitterbug was originally designed to infer network
congestion in access networks. Our work, on the other hand, is
focused on probe selection after identifying similarities in
\textit{all} latency anomalies across devices, which may not
necessarily be caused by congestion. We detail these modifications in
Section~\ref{sec:latency_elevation_detection}.

\fb{After reading the modifications below, I wonder whether it makes
  sense to summarize Jitterbug's limitations here before jumping into
  how we modify things. The modifications all make sense, they just
feel "sparse". Adding this summary upfront helps guiding the reader.}
\paul{Do we want to call this section ``Jitterbug'' or something more
general, since we are not only describing Jitterbug but also our modifications?}

\subsection{Probe Selection}
Selection of representative probes in active Internet measurement has
been extensively studied, with strategies ranging from improving
geographic footprint to more sophisticated methods that also take
coverage into account. Akella and Seshan \cite{akella2003empirical}
show that using only academic testbeds like PlanetLab can bias
results, and propose selecting probes based on traffic patterns and
popular destinations. Barford \textit{et al.}
\cite{barford2001marginal} find that adding more probes has
diminishing returns, suggesting that a small set of well-placed
probes can provide autonomous system (AS)-level visibility. Recent work focuses on
maximizing coverage and diversity: Holterbach \textit{et al.}
\cite{holterbach2017measurement} proposed selecting topologically
dissimilar probes, and the Metis algorithm improves RIPE Atlas
coverage by targeting underrepresented ASes and regions
\cite{appel2022metis, bajpai2015lessons}. Others use BGP or topology
data to choose the minimal set of probes that observe the most paths
\cite{shavitt2009quantifying}. Our work differs from these studies by
relying on no prior knowledge of the network topology. We use purely
an end-to-end, data-driven approach to identify a minimal set of
probes that can cover shared latency anomalies.

\subsection{Maximum Weighted Set Coverage Problem}

The maximum weighted set coverage problem is a widely studied
combinatorial optimization problem \cite{nemhauser1978analysis,
khuller1999budgeted}. Given a set of items $U$ and a collection of
subsets $S_1, S_2, \ldots, S_n$ such that $S_i \subseteq U$, the goal
is to select a subset of these sets such that the union of the
selected sets covers as many items in $U$ as possible. Each set $S_i$
has an associated weight $w_i$, and the objective is to maximize the
total weight of the selected sets while ensuring that each item in
$U$ is covered at least once.

An important property of this problem is that it is NP-hard, implying
that there is no known polynomial-time algorithm that guarantees an
optimal solution. However, there are several approximation algorithms
that provide close-to-optimal solutions in polynomial time. One of
the most common approaches is the greedy approach, which iteratively
selects the set that covers the largest additional weight of
uncovered elements at each step. This approach continues until the
desired number of sets (up to a given budget $k$) is selected.
Nemhauser et al.~\cite{nemhauser1978analysis} showed that this greedy
algorithm achieves a $(1 - 1/e) \approx 63\%$ approximation to the
optimal solution, which is the best possible guarantee unless $P =
NP$. In Section~\ref{sec:algorithm}, we propose a similar greedy
algorithm for selecting representative probes in our deployment.

This problem has applications in various fields, including outbreak
detection \cite{leskovec2007cost}, sensor placement
\cite{krause2008near}, social network analysis
\cite{kempe2003maximizing}, and text summarization
\cite{lin2011class}. In our work, we reformulate the problem of
selecting representative devices as an instance of this problem,
where the items in each set are latency anomalies, which may or may
not be shared across devices. The weights of the sets are determined
by an ``impact" metric, which is a function of the amplitude and
duration of the shared anomalies. The goal is to select a subset of
devices that maximizes the total impact while minimizing the number
of unique anomalies detected in the selected devices. We posit that
the definition of impact, albeit simple in our case, can be extended
to more nuanced definitions in future work. For example, one could
consider defining impact in terms of QoE degradations
\cite{mangla2019using,michel2022enabling,sharma2023estimating} for a
particular application, or in terms of the geographic area of
occurrence for a particular latency anomaly. We omit these
explorations in our current work, but we believe they are important
avenues for future research.

\section{Dataset} \label{sec:methods}

In this section, we outline our data collection approach and share
basic descriptive statistics of our dataset. We conclude with a
discussion on the capability of our dataset to capture shared network anomalies.

\subsection{Data Collection and Measurement Frequency}
\label{sec:data_collection}

We collect the data for this study primarily from residential
networks using an open-source measurement platform\footnote{We
withhold the platform name to preserve our submission's anonymity.}.
The platform allows for the collection of a wide suite of network
performance tests such as throughput, latency, application Quality of
Experience (QoE) and network path metrics. Our platform is typically
deployed on a Raspberry Pi 4B (RPi) probe that connects directly to
the home router using an Ethernet cable. The probe runs a set of
tests periodically to measure various network performance metrics. \paul{As Francesco mentioned in slack, we might
want to head-off any RPi complaints here.} While RPis are
resource-constrained devices, prior benchmarking work
\cite{papakyriakou2023benchmarking} has shown the RPi 4B to achieve
gigabit speeds on wired connections, which is suitable for running
network performance tests. Further, our focus in this work is on idle
latency measurements, which are not significantly affected by the
RPi's performance. We also note that our measurement suite can be
easily adapted to run on other platforms.

\begin{table}[htbp]
  \centering
  \begin{minipage}{0.48\linewidth}
    \centering
    \captionsetup{width=.77\linewidth}
    \begin{tabular}{ll}
      \toprule
      \textbf{Destination} & \textbf{IP Address} \\
      \hline
      Atlanta & \texttt{4.71.254.129} \\
      \rowcolor{lightgray} Chicago & \texttt{4.71.251.129} \\
      Denver & \texttt{4.34.58.1} \\
      \rowcolor{lightgray} Johannesburg & \texttt{196.24.45.129} \\
      Paris & \texttt{77.67.119.129} \\
      \rowcolor{lightgray} Seattle & \texttt{38.102.0.65} \\
      Stockholm & \texttt{195.89.146.193} \\
      \rowcolor{lightgray} Tunis & \texttt{41.231.21.1} \\
      \bottomrule
    \end{tabular}
    \caption{Destinations used for latency measurements.}
    \label{tab:ping_destinations}
  \end{minipage}\hfill
  \begin{minipage}{0.48\linewidth}
    \centering
    \begin{tabular}{lr}
      \toprule
      \textbf{ISP} & \textbf{Number of devices} \\
      \midrule
      Comcast & 62 \\
      \rowcolor{lightgray} AT\&T & 29 \\
      RCN & 4 \\
      \rowcolor{lightgray} Everywhere Wireless & 2 \\
      Webpass & 2 \\
      \rowcolor{lightgray} Campus Network & 1 \\
      Verizon & 1 \\
      \bottomrule
    \end{tabular}
    \caption{Distribution of the number of devices by ISP.}
    \label{tab:isp-distribution}
  \end{minipage}
\end{table}

We leverage latency measurements from our platform to identify the
presence of shared network anomalies among 99 user devices deployed
between April 2022 and July 2022. Although our devices are deployed
even during the present day, we use this time period because it is when most of our devices were actively collecting data.
Each probe reports a new ICMP round-trip latency measurement every five
minutes, and performs one NDT and one Ookla speed test every hour.
We primarily use latency measurements destined to M-Lab servers as
these provide a global coverage along with being destinations with
fixed IP addresses. Table~\ref{tab:ping_destinations} shows
the list of destinations used for latency measurements that we
analyze for this work. The destinations are chosen to be
geographically diverse, with a mix of locations in North America,
Europe, and Africa, to guarantee topological diversity. Each selected server is
located outside the AS of the probe's ISP to ensure that we are
measuring the end-to-end path and not just the last-mile or middle-mile.

\subsection{Dataset Overview} \label{sec:dataset_overview}

\begin{figure*}[t!]
  \centering
  \hfill
  \begin{subfigure}[t]{0.3\linewidth}
    \centering
    \captionsetup{width=.85\linewidth}
    \includegraphics[width=\linewidth]{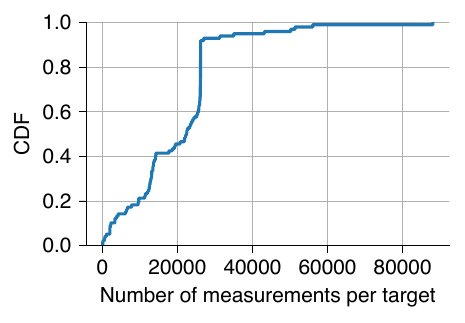}
    \caption{Measurement count per probe for last-mile latency.}
    \label{fig:num-meas-cdf}
  \end{subfigure}
  \hfill
  \begin{subfigure}[t]{0.3\linewidth}
    \centering
    \captionsetup{width=.85\linewidth}
    \includegraphics[width=\linewidth]{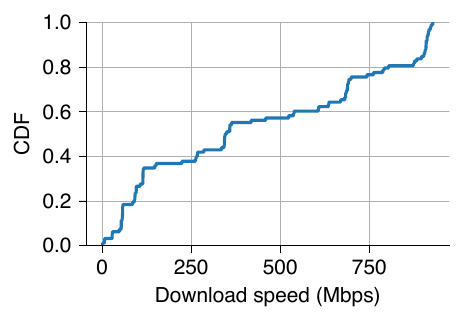}
    \caption{Median download speed distribution per probe.}
    \label{fig:download-speed-cdf}
  \end{subfigure}
  \hfill
  \begin{subfigure}[t]{0.3\linewidth}
    \centering
    \captionsetup{width=.95\linewidth}
    \includegraphics[width=\linewidth]{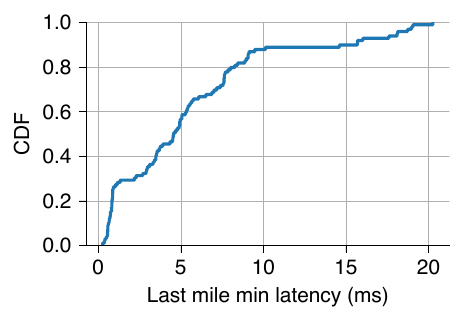}
    \caption{Minimum last-mile latency distribution per probe.}
    \label{fig:last-mile-latency-cdf}
  \end{subfigure}
  \caption{\takeaway{An overview of basic descriptives of our dataset
  of 99 devices.}}
  \label{fig:descriptives}
\end{figure*}

Our deployment consists of 99 Raspberry Pi devices located in the
city of Chicago. \fb{Are zip code the real ones? If they are, there
  is no point in hiding the city. Otherwise we should mask the zip
codes}\paul{I'd vote not anonymizing the city. The university .edu
sure, but this is an open dataset} These are distributed across 31
distinct zip codes, with the majority concentrated in just two. The
remaining zip codes each contain between one and six devices, and we
use data from these devices to validate our findings. A
well-documented contrast in terms of social demographics between the
two zip codes motivated this sampling decision~\cite{kaufman2013chicago}. Despite this
heterogeneity, we do not constrain our methods to look for shared
events between these two zip codes, as we expect anomalies to be
shared across zip codes as well, especially if they originate at the
middle-mile or beyond.

Table~\ref{tab:isp-distribution} shows the distribution of probe
counts by ISP. The devices are connected to seven distinct ISPs, but
our sample is dominated by Comcast and AT\&T, which are the two most
prominent ISPs in the area. The remaining ISPs are either smaller or
provide specialized services, such as campus networks or fixed
wireless access. In Section~\ref{sec:shared_elevations}, we analyze
the impact of devices being in the same ISP on shared network anomalies.

Figure~\ref{fig:num-meas-cdf} shows the distribution of the number of
latency measurements per probe. We use the last-mile latency
measurement counts to construct this plot, and expect similar numbers
for other targets as well. We see that more than 78\% of devices have
more than 10,000 measurements, with a median of 22,454 measurements
per probe, per target. Additionally, we notice that two devices
contain only 171 and 108 measurements, respectively, and do not
register any anomalies within less than one day of deployment. We
exclude these devices from our analysis, making our final dataset
size 97 devices. Overall, the remaining devices offer a rich set of
measurements, all of which may not perfectly align in time, but are
still useful for our analysis.

Next, we assess typical speeds for each probe based on NDT speed test
measurements, which we conduct at an hourly frequency.
Figure~\ref{fig:download-speed-cdf} shows the distribution of median
download speed across devices. We observe a step-like distribution,
with each step likely signifying distinct speed tiers, similar to
distinct clusters observed in crowdsourced datasets in prior work
\cite{paul2022importance}. About 7\% of devices have a median
download speed of less than 50 Mbps, over 43\% above 500 Mbps and
nearly 15\% above 900 Mbps, suggesting a good mix of devices with
different speed tiers. The aggregate median download speed across all
devices is about 347 Mbps, which is close to the typical speed
observed by the FCC MBA program
\cite{fccmba2025} around the same time period as our deployment
\cite{fcc2023fixed}. Overall, we observe a range of speeds across our
devices, suggesting that our analysis is not biased towards any
particular speed tier.

We also look at the minimum last-mile latency distribution for the
devices. We use the first public hop outside the home network as the hop for measuring
last mile latency. Our last mile test involves sending traceroute
probes to a fixed target, followed by grabbing the first public hop
from the output. Then, we measure the round-trip time to this hop
using ICMP echo requests. Figure~\ref{fig:last-mile-latency-cdf}
displays the distribution of minimum last mile latency across
devices. We observe that a majority of devices (> 80\%) have a last
mile latency of less than 10 ms, which is indicative of fiber and
cable connections. Nearly 10\% of devices have a last mile latency of
10-20 ms, which is due to the presence of DSL connections in the
sample; over 15 devices showed download speeds below 50 Mbps.
Overall, our dataset contains a diverse set of devices with a mix of
access technologies that we typically observe in residential networks.

While our sample is not representative of the entire U.S. population,
it does provide a diverse set of devices connected to different ISPs,
speed tiers and access technology types in an urban area. This
diversity allows us to capture a broad range of real-world network
conditions and performance behaviors, making our dataset suitable for
analysis of shared network anomalies.

\fb{I guess the description of the dataset is not complete}

\begin{figure*}[t!]
  \centering
  \includegraphics[width=0.75\textwidth]{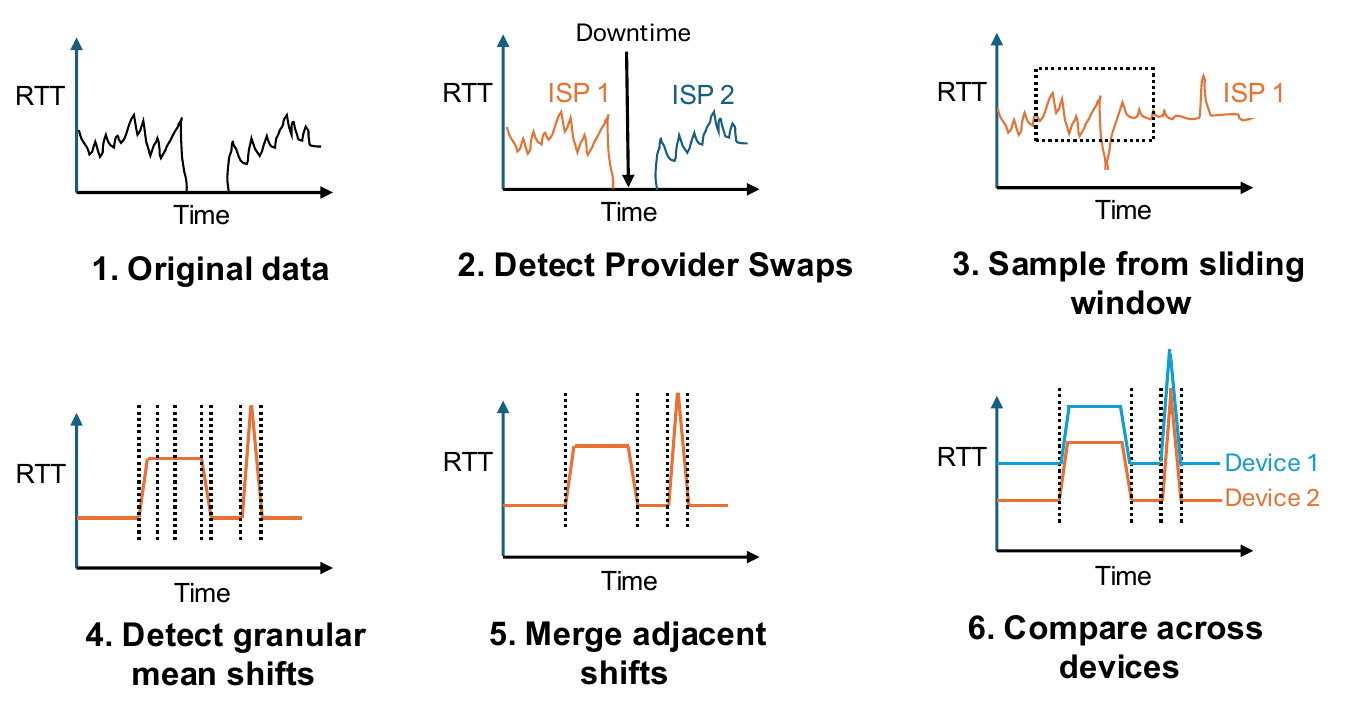}
  \caption{An overview of our anomaly detection methodology. We
    detect mean shifts with sensitive parameters for every provider
    change in the dataset. After merging adjacent shifts, we analyze
    the co-occurrence of detected anomalies across devices located in
  the same geography.}
  \label{fig:overview}
\end{figure*}

\section{Spatio-temporally Correlated Latency Anomalies}
\label{sec:spike-behavior}

\fb{I would clearly state the goals of the section (and possibly a
quick summary of the results) in the beginning.}
\paul{Minor: this section title and the next are both sort of vague /
  dry. I wish I had a good suggestion off the top of my head but we
want to succinctly convey what the reader is about to read.}

In this section, we present our methodology for detecting shared
latency anomalies in time and space. We begin by describing our
approach to detect latency anomalies using a modified version of
the Jitterbug congestion inference framework~\cite{carisimo2022jitterbug}. We then analyze the temporal
overlap of shared events of elevated latency between probe pairs, and
characterize the relationship between the similarity in amplitude and
the temporal overlap of these events. Finally, we assess the impact
of shared events of elevated latency.
\subsection{Anomaly Detection} \label{sec:latency_elevation_detection}

\fb{MINOR: I would move from textbf to a modified paragraph style
that removes indentation. Happy to implement this quickly if you want :).}

\paragraph{Extending Jitterbug with PELT.} We make a number of design
changes to the original Jitterbug methodology for our use case.
First, the authors of the original work report that the BCP algorithm
takes a longer time to process large datasets (60-90 seconds for 15
days of data). Since our dataset contains a total of 14.6 million
latency measurements from over 4 months of deployment, we extend the
Jitterbug pipeline to support a more lightweight yet effective
algorithm, called Pruned Exact Linear Time (PELT)
\cite{killick2012optimal}. PELT is based on a linear time dynamic
programming algorithm that detects multiple change-points in a
time-series by minimizing a cost function. It uses a penalty
parameter to control the sensitivity of the algorithm. A higher
penalty leads to the detection of fewer change-points. Since our
objective is to detect overlaps, we set this parameter to 0.001 ms to
detect maximum, albeit insignificant, changes in latency. This
results in a highly fragmented set of change-points, which we later
merge into a single jump if they are sufficiently close in time.

\paragraph{Modification of jump detection heuristics.} The second
change we make is to modify the heuristics used to detect jumps in
latency. While continuing with Jitterbug's original threshold of 0.5
ms to infer a mean shift, we notice that the original approach infers
both these cases as jumps: (1) a legitimate jump in latency from
baseline, and (2) recovery to baseline after a ``dip" in latency. We
take a number of steps to avoid this behavior. First, we call a given
segment a dip if its mean is lower than the mode of the entire
signal (our measure of baseline). Second, we track the index of both
the last detected dip and the last detected jump. Finally, we mark a
segment as a jump only if: (1) its mean is 0.5 ms higher than the
previous segment, (2) its mean is higher than the baseline, and (3)
the last segment was not a dip. On visual inspection, we find that
this helps significantly reduce the number of false positives in our dataset.

\paragraph{Modification to memory rule.} The third and final change
we make is to modify the memory rule Jitterbug uses to avoid
labelling the offset of a jump, i.e., the post-jump period of time
when latency is on its way back to the baseline, as a dip. The
original rule states that a change-point $C_2$ is also labelled as a
jump if the previous change-point $C_1$ is a jump and the mean
latency of $C_2$ is at least as large as that of $C_1$. Our
modification is to check whether the absolute difference in maximum
latency of adjacent segments is within 1.5 times the standard
deviation of the entire signal, and label the second segment as a
jump if the first segment is also a jump. This allows for a more
accurate re-labelling of offsets as jumps in our experiments.

\paragraph{Overview of the pipeline.} Our extensions to Jitterbug and
the resulting pipeline are summarized in Figure~\ref{fig:overview}.
As in the original implementation, we first convert the raw RTT time series to a
min-RTT series, with the min calculated in 15-minute bins. Then, we
look for changes in the ISP of the probe based on the WHOIS lookups
of the IP address used for each measurement. This is done to ensure
that we are only looking at latency changes that are likely to be
caused by a network event and not by a change enforced by the user.
We find that four users switched their ISP without notification during
our deployment, and we take measures to re-label and segment their
measurements accordingly. Next, we sample the min-RTT series using
sliding time windows with length 48 hours. We move the window by 24
hours at a time to ensure that the detected jumps are not located at
the edges of the window. Using larger windows to detect long-term
jumps is a ripe avenue for future work. In steps 4 and 5, we apply
our modified Jitterbug methodology to detect anomalies in latency. In
the final step, we compare these jumps across devices to compute
overlaps in time. We deem two anomalies $E_1(s_1, e_1)$ and $E_2(s_2,
e_2)$ to be overlapping if $s_1 \leq e_2$ and $s_2 \leq e_1$, where
$s_i$ and $e_i$ are the start and end times of the anomalies,
respectively. These overlaps are calculated uniquely at the level of
individual device pairs and the latency destination.

\fb{I am not sure of this, but worth a discussion: A reviewer might
question whether our modifications ``break" Jitterbug's original
accuracy. I see two ways of solving this: either state CLEARLY
upfront that Jitterbug does not work for our scenario; or show
through some sort of validation here.}

\begin{figure}
\centering
\includegraphics[width=0.45\linewidth]{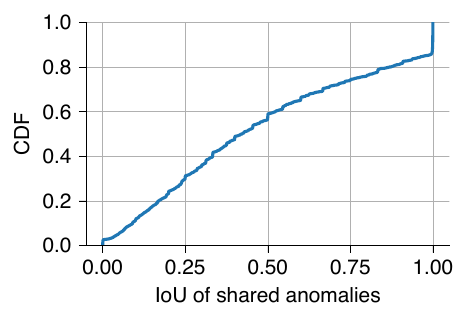}
\caption{A distribution of the IoU of shared events of elevated
  latency. \takeaway{Over 14\% of overlapping events exhibit an IoU of
0.99 or higher.}}
\label{fig:iou-distro}
\end{figure}

\subsection{Analysis of Shared Anomalies} \label{sec:shared_elevations}

In this section, we characterize the overlap of shared events of
elevated latency between probe pairs for a given destination across
time and space. Our goal is to determine how often these events align
in terms of magnitude across users in the same geography and the
extent to which they coincide in time. Additionally, we assess the
impact of these shared events across different levels of temporal
overlap. \paul{I might have missed it but I feel like we breeze past
the fact that this is spatio-temporal. We just dive into temporal and
I guess we rely on using the same zip code, but we should mention
that the goal includes the notion of space.}

\paragraph{Temporal overlap quantification.} To quantify the extent
to which anomalies are shared over time, we compute the intersection
over union (IoU) based on their durations. For anomalies $E_1(s_1,
e_1)$ and $E_2(s_2, e_2)$, the IoU is defined as:

\[
IoU(E_1, E_2) = \frac{max(0, min(e_1, e_2) - max(s_1, s_2))}{max(e_1,
e_2) - min(s_1, s_2)}
\]

Here $s_i$ and $e_i$ are the start and end times of $E_i$,
respectively. The IoU value ranges from 0 to 1, where 0 indicates no
overlap and 1 indicates complete overlap. We compute the IoU for each
pair of anomalies observed toward the same destination by a pair of devices.

\paragraph{Shared anomalies are not random.} We begin by analyzing
the distribution of IoU of shared events of elevated latency.
Figure~\ref{fig:iou-distro} shows a CDF of the IoU values observed
over a dataset of over 1.5 million paired events of elevated latency
toward M-Lab servers. We observe that over 23\% of overlapping events
show an IoU of 0.8 or higher, and over 14\% of overlapping events
exhibit an IoU of 0.99 or higher. To validate that this behavior is not
an artifact of the stochastic nature of network data or imperfect
change-point detection, we conduct a randomization test. Specifically, we randomly shuffle the timestamps of the shared events
within a day of their occurrence individually for each probe while
preserving their duration and amplitude. We then recompute the IoU
for the shuffled events for 1000 repetitions of the randomization to
obtain a null distribution over the IoU values. In the null
distribution, we observe only 1.458\% $\pm$ 0.009\% of overlapping
events to show an IoU of 0.8 or higher. This indicates that the
observed IoU values are significantly higher than what would be
expected by chance, confirming that a significant number of events of
elevated latency are correlated in time.

\paragraph{Measuring similarity in amplitude.} Next, we examine the
relationship between the similarity in amplitude and the temporal
overlap for shared latency events. We compute the amplitude as the
difference between the maximum latency during an anomaly and the
baseline latency. Under the assumption that latency anomalies are
rare events, we use the mode of the minimum latency (binned in
15-minute intervals) as the baseline latency. Duration, on the other
hand, is defined as the difference in hours between the start and end
times of the anomaly. For a pair of shared events of elevated
latency, we define \textit{amplitude similarity} as the ratio between
the minimum and maximum amplitude of the shared events.

\begin{figure*}[t!]
\centering
\hfill
\begin{subfigure}[t]{0.3\linewidth}
  \centering
  \includegraphics[width=\linewidth]{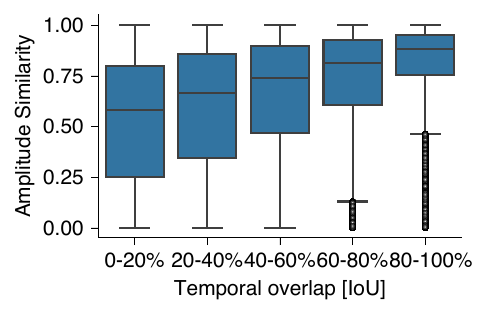}
  \caption{Overall amplitude similarity relationship with IoU.}
  \label{fig:amp-sim-overall}
\end{subfigure}
\hfill
\begin{subfigure}[t]{0.3\linewidth}
  \centering
  \includegraphics[width=\linewidth]{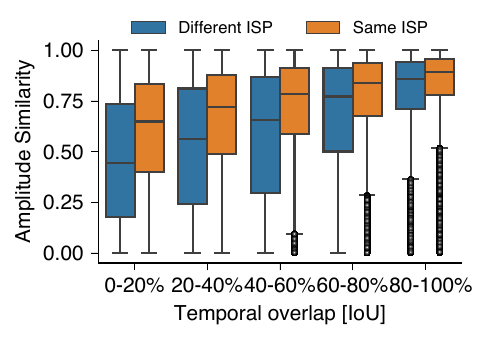}
  \caption{Amplitude similarity versus IoU stratified by shared or
  non-shared ISP.}
  \label{fig:amp-sim-isp}
\end{subfigure}
\hfill
\begin{subfigure}[t]{0.3\linewidth}
  \centering
  \includegraphics[width=\linewidth]{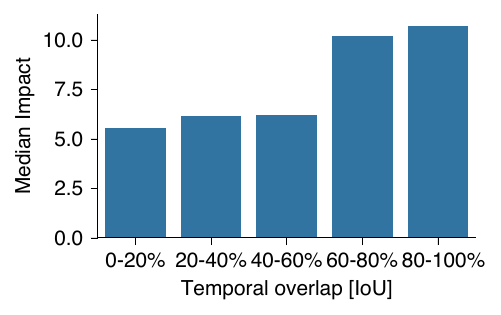}
  \caption{Median impact (ms-hours) versus IoU.}
  \label{fig:impact-iou}
\end{subfigure}
\caption{\takeaway{Events with similar amplitudes often exhibit
    greater overlap in duration, indicating they likely reflect the same
    underlying phenomenon. This relationship is more pronounced when the
    events occur within the same ISP. Moreover, events with higher
temporal overlap (IoU) tend to be the ones with greater impact.}}
\label{fig:impact-sim}
\end{figure*}

\paragraph{Temporally coincident anomalies have similar amplitudes.}
Figure~\ref{fig:amp-sim-overall} shows the overall amplitude
similarity relationship with IoU. We observe that the similarity in
amplitude increases with the IoU of the shared anomalies, with a
median amplitude similarity of 0.88 for the 80-100\% IoU range. This
indicates that events with a high temporal overlap are also more
likely to be similar in terms of their amplitude. These findings
suggest that temporal alignment can be a strong indicator of
underlying shared causes across measurement points, and that
amplitude similarity can be used to attributing events to a common
root cause. We also stratify the events by whether they are shared by
the same ISP or not. Figure~\ref{fig:amp-sim-isp} shows the amplitude
similarity versus IoU for events shared by the same ISP and those that
are not. We observe that the correlation between amplitude similarity
and IoU is slightly stronger for events shared by the same ISP, with
a median amplitude similarity of 0.89 for the 80-100\% IoU range,
compared to 0.85 between ISPs. This indicates that events shared by
the same ISP are slightly more likely to be similar in terms of their
impact. We note that the overall relationship between amplitude
similarity and the IoU is not perfectly linear (Spearman's
correlation coefficient = 0.37) due to the presence of outliers in
the data. Since our change-point detection approach is applied
individually to each probe, it is possible to get different amplitude
values and less-than-perfect IoUs for the same event across devices
due to noise in the data.

\paragraph{High impact anomalies exhibit greater overlap.}  We next
assess the IoU ranges that are most likely to be associated with a
high impact. We define the \textit{impact} of events of elevated
latency (in units of ms-hours) as the product of their amplitude and
duration. We compute the median impact for each 20\% IoU bin.
Figure~\ref{fig:impact-iou} shows the median impact (ms-hours) of
shared events of elevated latency as a function of the 20\% IoU bin
they belong to. We observe that the median impact of shared events
increases with the IoU, with the 60-100\% IoU ranges showing a median
impact of nearly 10 ms-hours. We note that a 10 ms-hour impact
corresponds to the 72-percentile of the overall impact distribution,
indicating that events showing a high mutual overlap tend to be more
impactful than a typical event of elevated latency.

\begin{figure}
\centering
\begin{subfigure}[t]{0.48\linewidth}
  \centering
  \includegraphics[width=\linewidth]{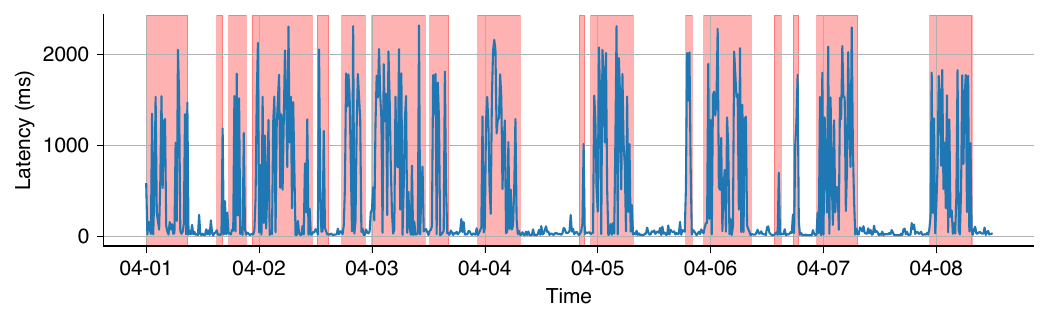}
  \caption{A likely case of bufferbloat; our approach detects a
  large amplitude for the anomaly.}
  \label{fig:likely-bufferbloat}
\end{subfigure}
\hfill
\begin{subfigure}[t]{0.48\linewidth}
  \centering
  \includegraphics[width=\linewidth]{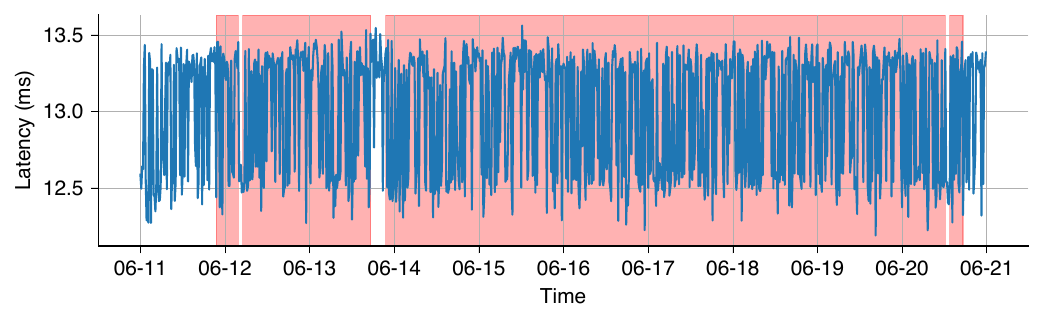}
  \caption{A case with high variation in baseline latency; our
  approach leads to overestimation of duration.}
  \label{fig:high-variation}
\end{subfigure}
\caption{Examples of outliers in impact values. Red shaded regions
denote detected anomalies.}
\label{fig:outliers}
\end{figure}

\paragraph{Impact of outliers.} It is worth noting that outliers in
both amplitude similarity and impact values can potentially affect
our analysis. Some reasons for such scenarios could be merging of
multiple tiny jumps in latency into a single large anomaly,
bufferbloat, or due to a malfunctioning probe/home router. An example
of two such cases is shown in Figure~\ref{fig:outliers}. In
Figure~\ref{fig:likely-bufferbloat}, we observe a large amplitude for
the anomaly, which is likely due to bufferbloat or a malfunctioning
probe. We also observe that this probe shows an atypical distribution
of overall IoU values, with only 5\% of the shared events showing an
IoU of 0.8 or higher. In Figure~\ref{fig:high-variation}, we observe
a case with oscillatory behavior in baseline latency between the
12.5-13.5 ms range. Since our methodology involves marking any jump
with a higher than 0.5 ms mean shift as an anomaly, we observe many small jumps in latency being merged into a single
large anomaly during the de-duplication process. This results in a
large impact value of 168.1 ms-hours for the anomaly between June 13
and June 21. This particular anomaly shows a maximum IoU score of
only 0.52 across all its shared events because of its large duration,
suggesting that most of the overlaps associated with this probe are
spurious. In Section~\ref{sec:algorithm}, we discuss how our greedy
algorithm for probe selection can be made robust to such outliers.

\section{Data Reduction through Probe Selection}
\label{sec:probe-selection}

In this section, we propose an algorithm to select a subset of probes
that maximizes the coverage of high-impact latency anomalies across
the dataset. Our ultimate goal is to reduce measurement cost and
redundancy while ensuring that we cover the most impactful
network events. We first formulate the problem as a maximum weighted
set coverage problem, which is known to be NP-hard. We then present a
greedy approximation algorithm with provable guarantees, followed by
an empirical evaluation on our dataset.

\subsection{Problem Formulation} \label{sec:problem-formulation}

Let $\mathcal{P}$ denote the set of all probes deployed, and let
$\mathcal{E}$ be the set of high-impact latency anomalies observed in
the dataset. Each anomaly $e_k \in \mathcal{E}$ is associated with an
impact score $I_k$, defined as the product of its amplitude and
duration. Let $\mathcal{E}(p_i) \subseteq \mathcal{E}$ denote the set
of anomalies observed by probe $p_i \in \mathcal{P}$.

The goal is to find the smallest subset of probes $\mathcal{P}'
\subseteq \mathcal{P}$ such that the total impact of the anomalies
they collectively observe satisfies:

\[
  \sum_{e_k \in \bigcup_{p_i \in \mathcal{P}'} \mathcal{E}(p_i)} I_k
  \geq c \cdot \sum_{e_k \in \mathcal{E}} I_k
\]

where $c \in (0, 1]$ is a user-defined coverage threshold (e.g., $c =
0.65$ for 65\% impact coverage) to control the trade-off between
minimizing the number of probes and maximizing coverage of impactful
events. Since the absolute impact values $I_k$ are subject to noise
from change-point detection, it is likely that the impact of selected
probes may exceed the coverage threshold with only a few selected
probes. We verify this using pilot experiments, where using the
absolute impact values led to overestimation of coverage using a
limited number of probes. To minimize this effect, we replace $I_k$ with its
log-transform, $\log(1 + I_k)$ in the above inequality. This transformation shrinks the original range of impact values, making probe selection less sensitive to noise from change point detection while
preserving the relative ordering of anomalies. We call this
value \textit{log-impact}.

\subsection{Algorithm \& Implementation} \label{sec:algorithm}

\paragraph{Uniquely identifying latency anomalies.} Before designing
an algorithm to select the minimum number of probes, it is critical
to assign a unique identifier to each anomaly $e_k$ in the dataset.
This helps us determine the set of probes that observe an anomaly due
to the same network perturbation. Guided by previous discussions, we
assign the same unique identifier to a pair of anomalies if they
share the same destination and overlap significantly in time (IoU
$\geq \delta_{IoU}$). It is important to note that the choice of
$\delta_{IoU}$ can significantly affect the total number of unique
identifiers assigned to the anomalies. More relaxed values lead to
more unique identifiers, which ultimately increases the total number
of probes needed to cover the same set of anomalies. Further, the
choice of change-point detection approach can also affect the number
of unique identifiers assigned to the anomalies as there could be
imperfect durational overlaps between devices. In
Section~\ref{sec:eval-algorithm}, we experiment with different values
of $\delta_{IoU}$ and the coverage fraction $c$, to find the best
trade-off between minimizing the number of probes and maximizing
coverage of impactful events. We originally considered using
amplitude similarity ($\delta_{sim}$) as an
additional parameter for assigning unique identifiers, but chose to
exclude it for two reasons: (1) our analysis visualized in
Figure~\ref{fig:amp-sim-isp} suggests that barring few outliers,
temporal overlap correlates with amplitude similarity and (2)
relying on a single parameter reduces complexity and can provide better
explainability in auditing our algorithm.

\paragraph{Greedy Heuristic.} Our algorithm iteratively selects the
probe that maximizes the marginal gain in coverage until the desired
coverage threshold is met. Algorithm~\ref{alg:greedy-probe-selection}
outlines this approach. The algorithm maintains a set of selected
probes $\mathcal{P}'$ and a set of covered anomalies $\mathcal{S}$.
In each iteration, it selects the probe $p^*$ that maximizes the sum
of log-impact scores of the anomalies that are not already covered by
the selected probes. The algorithm continues until the cumulative
log-impact $C$ of the covered anomalies reaches a fraction $c$ of the
total log-impact $T$.

\begin{algorithm}
  \caption{Greedy Probe Selection Algorithm}
  \label{alg:greedy-probe-selection}
  \begin{algorithmic}[1]
    \State $\mathcal{P}' \gets \emptyset$ \Comment{Selected probe set}
    \State $\mathcal{S} \gets \emptyset$ \Comment{Set of covered anomalies}
    \State $C \gets 0$ \Comment{Cumulative log-impact}
    \State $T \gets \sum_{e_k \in \mathcal{E}} \log(1 + I_k)$
    \Comment{Total log-impact to cover}
    \While{$C < c \cdot T$}
    \State $p^* \gets \argmax_{p_i \in \mathcal{P} \setminus
    \mathcal{P}'} \sum_{e_k \in \mathcal{E}(p_i) \setminus
    \mathcal{S}} \log(1 + I_k)$
    \State $\mathcal{P}' \gets \mathcal{P}' \cup \{p^*\}$
    \State $\mathcal{S} \gets \mathcal{S} \cup \mathcal{E}(p^*)$
    \State $C \gets \sum_{e_k \in \mathcal{S}} \log(1 + I_k)$
    \EndWhile
    \State \textbf{return} $\mathcal{P}'$
  \end{algorithmic}
\end{algorithm}

The overall time complexity of this algorithm is $O(n^2 \cdot m)$, where $n$ is the number of probes and $m$ is the number of unique anomalies. We believe this is acceptable for most practical applications. This problem formulation and algorithm can be proven to achieve a
guaranteed approximation of the optimal solution for a given probe
budget. We refer an interested reader to relevant literature for more
details on the exact approximation ratio and proof \cite{nemhauser1978analysis}.

\subsection{Empirical Evaluation} \label{sec:eval-algorithm}

We now empirically evaluate the performance of our probe selection
algorithm on latency measurements from the dataset described in
Section~\ref{sec:data_collection}. First, we evaluate the
relationship between choices of the coverage fraction $c$, and the
total number of probes selected with fixed values of $\delta_{IoU}$.
We then evaluate the effect of varying the threshold, $\delta_{IoU}$
on the number of probes selected. Finally, we compare our algorithm
with baseline approaches, such as uniform random sampling and a
na\"ive baseline that selects probes based on the descending order of
their impact scores.

\begin{figure}
  \centering
  \begin{subfigure}[t]{0.32\linewidth}
    \centering
    \includegraphics[width=\linewidth]{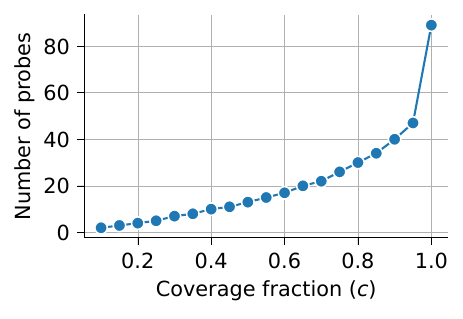}
    \caption{The number of selected probes (out of 97) increases
      steadily with increasing coverage fraction until $c = 0.95$, beyond
      which there are diminishing returns in coverage. The latency
    destination used for this plot is an M-Lab server in Chicago.}
    \label{fig:coverage-ratio}
  \end{subfigure}
  \hfill
  \begin{subfigure}[t]{0.32\linewidth}
    \centering
    \includegraphics[width=\linewidth]{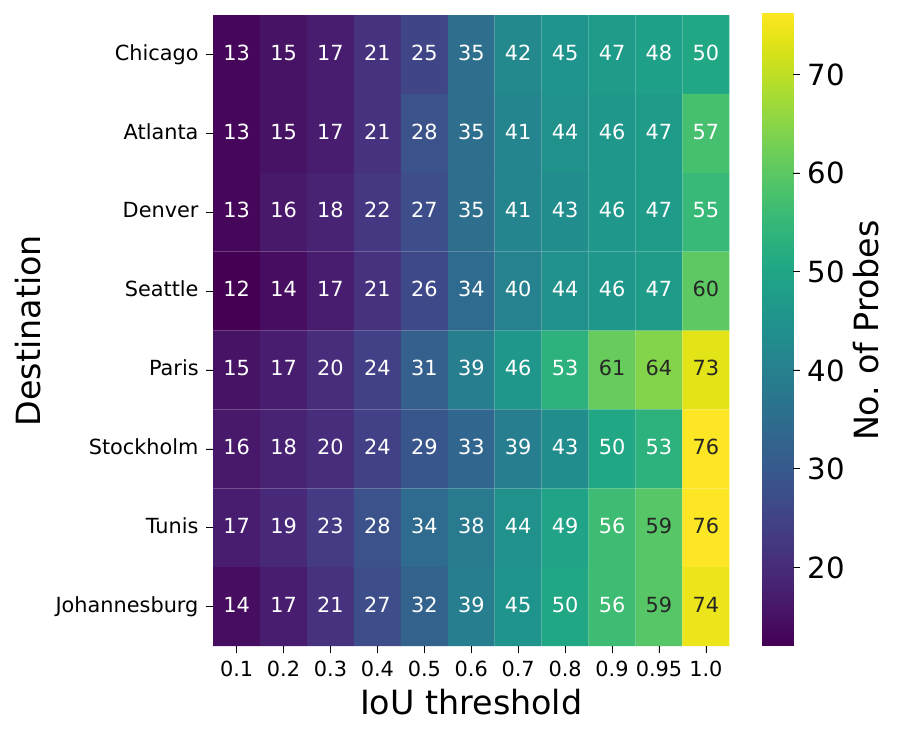}
    \caption{A heatmap showing the number of selected probes (out of
      97) as a function of the threshold $\delta_{IoU}$ across
      destinations. Destinations (top-bottom) are sorted by increasing
    great circle distance from Chicago.}
    \label{fig:probe-count-vs-thresholds}
  \end{subfigure}
  \hfill
  \begin{subfigure}[t]{0.32\linewidth}
    \centering
    \includegraphics[width=\linewidth]{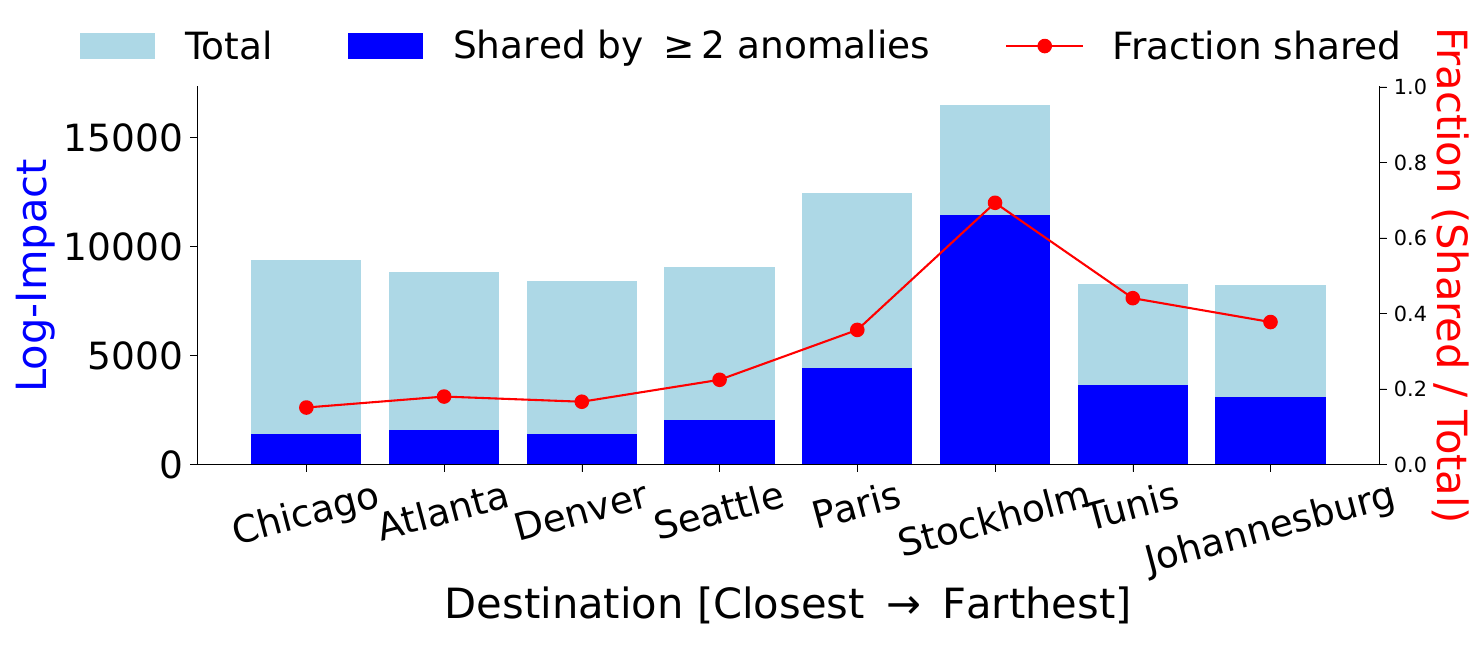}
    \caption{Comparison of total log-impact distribution across
      destinations with the log-impact observed for probes that share
    anomalies ($\delta_{IoU} = 0.9$).}
    \label{fig:impact-sharing}
  \end{subfigure}
  \caption{We require a greater number of probes to achieve 95\%
    impact coverage with higher values of $\delta_{IoU}$. The fraction
    of shared log-impact increases with distance until Stockholm, but
  drops for farther targets.}
  \label{fig:comparative-performance}
\end{figure}

\paragraph{Probe count vs. coverage fraction.} To evaluate the trade-off
between coverage fraction and the number of selected probes, we fix
$\delta_{IoU} = 0.9$ and vary the coverage fraction $c$ from 0.1 to 1.0
in increments of 0.05. The choice of $\delta_{IoU}$ is motivated by
our results from Figure~\ref{fig:impact-iou}, which shows that a
significant fraction of high impact anomalies have an IoU $\geq 0.6$.
We pick $\delta_{IoU} = 0.9$ as it ensures that there is a tight
overlap between anomalies detected by different probes.
Figure~\ref{fig:coverage-ratio} shows the number of probes selected
as a function of the coverage fraction. As expected, the number of
selected probes increases with increasing coverage fraction. Out of a
total of 97 probes used in the deployment, the number of selected
probes is 1 when $c = 0.1$, 47 when $c = 0.95$, and sharply increases
to 89 when $c = 1.0$. Only 89 probes are selected out of 97 when $c =
1.0$ because the algorithm is able to cover nearly all the impact of
latency anomalies, suggesting that the remaining probes do not
contribute any new anomalies based on the threshold used. The number
of probes selected increases slowly until $c = 0.95$, and provides
marginal gains in coverage after this point, suggesting that the
algorithm is able to cover a significant portion of the total impact
with less than half the probes. We observe a similar trend for other
latency destinations, with greater number of probes needed for
destinations farther away from the probe deployment location to
achieve 95\% impact coverage. This is likely because of longer paths
taken by packets to reach these destinations, leading to more diverse
network perturbations being observed.

\paragraph{Impact of temporal overlap threshold.}
To evaluate this effect, we fix the coverage fraction to 0.95 and vary the anomaly temporal overlap
threshold $\delta_{IoU}$ from 0.1 to 1.0 in increments of 0.1. For each
value, we run Algorithm~\ref{alg:greedy-probe-selection} and record the
number of probes required to meet the coverage criterion.
Figure~\ref{fig:probe-count-vs-thresholds} shows the resulting probe
counts across multiple targets. As $\delta_{IoU}$ increases, the number
of selected probes grows, reaching a maximum of 76 at $\delta_{IoU}=1.0$,
as stricter overlap reduces anomaly grouping and redundancy. Conversely,
relaxing the threshold to 0.1 reduces the probe count to 12 by
increasing overlap among similar events. We also observe significant destination-specific variation. This variation is driven by the fraction of log-impact shared across probes
and by the effect of distance on anomaly detectability.
Figure~\ref{fig:impact-sharing} shows that shared log-impact increases
with distance and peaks at Stockholm, likely due to common upstream
infrastructure. For more distant targets (e.g., Tunis and Johannesburg),
total detected impact decreases due to fewer detectable anomalies, as
larger baseline RTTs and path diversity make small latency increases
harder to detect, which leads us to select more probes for comparable coverage.

\begin{figure}
  \centering
  \begin{subfigure}[t]{0.48\linewidth}
    \centering
    \includegraphics[width=\linewidth]{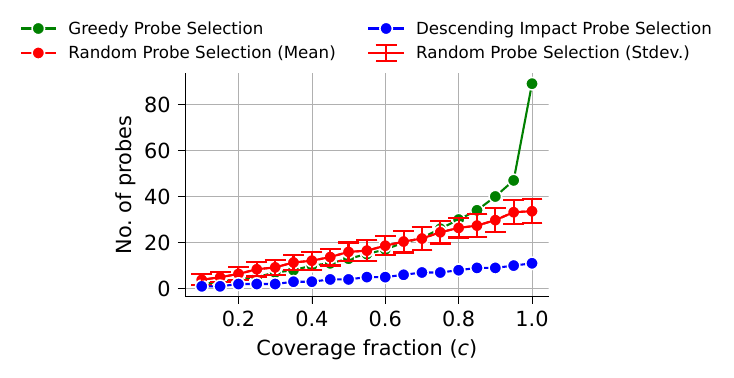}
    \caption{Probe counts for each selection method.}
    \label{fig:probe-counts}
  \end{subfigure}
  \hfill
  \begin{subfigure}[t]{0.48\linewidth}
    \centering
    \includegraphics[width=\linewidth]{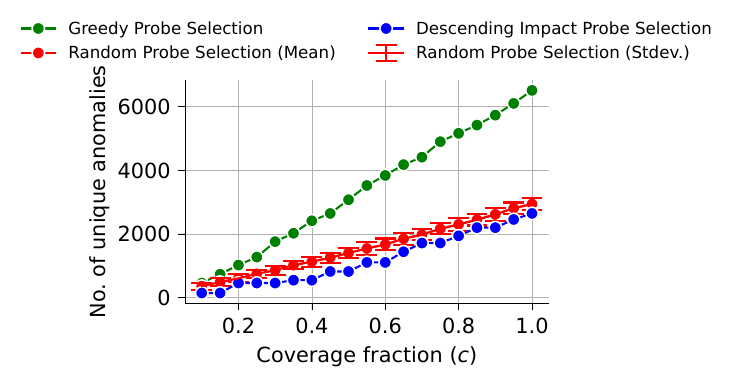}
    \caption{Number of unique anomalies observed by the selected
    probes across methods.}
    \label{fig:unique-spikes}
  \end{subfigure}
  \caption{A comparison of our approach with two baseline approaches.
    Our approach covers significantly higher number of unique anomalies
    compared to the baselines, while selecting a comparable number of
  probes to random sampling.}
  \label{fig:baseline-comparison}
\end{figure}

\paragraph{Comparison with baselines.}
We compare our probe selection algorithm against two baselines:
(1) uniform random sampling until the coverage threshold is met, and
(2) selecting probes in descending order of log-impact without
accounting for overlap. We evaluate performance using the number of
probes required to achieve the target coverage and by the number of
unique anomalies observed. We repeat random sampling 100 times and average
the results.
We evaluate all methods using a Chicago-based M-Lab server as the
destination and set $\delta_{IoU}=0.9$.
Figure~\ref{fig:baseline-comparison} summarizes the results, with error
bars indicating standard deviation for random sampling. Our approach
selects a similar number of probes as random sampling and more probes
than the descending log-impact baseline to achieve comparable coverage
(Figure~\ref{fig:probe-counts}), but consistently captures
significantly more unique anomalies across nearly all coverage levels
(Figure~\ref{fig:unique-spikes}). This indicates that our method better
prioritizes diversity in high-impact anomalies, whereas the baselines
tend to select redundant probes. This trend persists for geographically
farther targets.

\section{Predictive Evaluation} \label{sec:relevance}

Our approach thus far has focused on a greedy set cover based
post-hoc optimization of probe selection to reduce redundancy in
detected anomalies. We started with a non-random deployment of probes
in a single urban area and showed that we can reduce the number of
probes needed to capture a high fraction of the total anomaly impact.
This finding can readily allow for dimensionality reduction of
datasets collected from existing probe deployments that collect
end-to-end delay measurements (e.g., FCC MBA \cite{fccmba2025}, RIPE
Atlas \cite{ripeatlas2025}, and M-Lab \cite{mlab2025}). However, a
more interesting question is to ask: \textit{how well does the reduced set of probes predict future anomalies?}
In this section, we perform a predictive evaluation of our approach
to answer this question. We also consider analyzing the geographic
distribution of the reduced probe set in Appendix~\ref{sec:geo-dist}.

\begin{table}[t!]
  \centering
  \setlength{\belowrulesep}{1pt}
  \resizebox{1.0\textwidth}{!}{
    \begin{tabular}{c *{16}{c}}
      \toprule
      \multirow{4}{*}{\shortstack{\textbf{Selection
      Window}\\\textbf{(\# days)}}} &
      \multicolumn{16}{c}{\textbf{M-Lab Server Location (Increasing
      order of distance $\rightarrow$)}} \\
      \cmidrule(lr){2-17}
      & \multicolumn{2}{c}{\textbf{CHI}} &
      \multicolumn{2}{c}{\textbf{ATL}} &
      \multicolumn{2}{c}{\textbf{DEN}} &
      \multicolumn{2}{c}{\textbf{SEA}} &
      \multicolumn{2}{c}{\textbf{PAR}} &
      \multicolumn{2}{c}{\textbf{STO}} &
      \multicolumn{2}{c}{\textbf{TUN}} &
      \multicolumn{2}{c}{\textbf{JOH}} \\
      \cmidrule(lr){2-3} \cmidrule(lr){4-5} \cmidrule(lr){6-7}
      \cmidrule(lr){8-9} \cmidrule(lr){10-11} \cmidrule(lr){12-13}
      \cmidrule(lr){14-15} \cmidrule(lr){16-17}
      & R & C & R & C & R & C & R & C & R & C & R & C & R & C & R & C \\
      \midrule

      7  & \textbf{0.72} & 17 & \textbf{0.80} & 15 & \textbf{0.71} &
      16 & \textbf{0.87} & 14 & \textbf{0.91} & 19 & \textbf{0.94} &
      9  & \textbf{0.93} & 12 & \textbf{0.92} & 17 \\
      \rowcolor{gray!20}
      14 & \textbf{0.74} & 19 & \textbf{0.82} & 15 & \textbf{0.72} &
      16 & \textbf{0.86} & 15 & \textbf{0.89} & 19 & \textbf{0.96} &
      13 & \textbf{0.94} & 14 & \textbf{0.92} & 17 \\
      21 & \textbf{0.74} & 19 & \textbf{0.82} & 16 & \textbf{0.73} &
      17 & \textbf{0.88} & 16 & \textbf{0.87} & 19 & \textbf{0.97} &
      14 & \textbf{0.95} & 15 & \textbf{0.92} & 18 \\
      \rowcolor{gray!20}
      28 & \textbf{0.73} & 19 & \textbf{0.75} & 17 & \textbf{0.71} &
      17 & \textbf{0.85} & 17 & \textbf{0.85} & 19 & \textbf{0.96} &
      15 & \textbf{0.94} & 16 & \textbf{0.93} & 19 \\
      35 & \textbf{0.71} & 19 & \textbf{0.74} & 17 & \textbf{0.69} &
      17 & \textbf{0.81} & 17 & \textbf{0.84} & 19 & \textbf{0.96} &
      16 & \textbf{0.94} & 16 & \textbf{0.93} & 19 \\
      \rowcolor{gray!20}
      42 & \textbf{0.70} & 19 & \textbf{0.74} & 17 & \textbf{0.68} &
      17 & \textbf{0.76} & 17 & \textbf{0.83} & 20 & \textbf{0.97} &
      17 & \textbf{0.94} & 16 & \textbf{0.94} & 19 \\
      49 & \textbf{0.70} & 19 & \textbf{0.75} & 17 & \textbf{0.67} &
      17 & \textbf{0.75} & 16 & \textbf{0.80} & 20 & \textbf{0.96} &
      17 & \textbf{0.92} & 16 & \textbf{0.94} & 19 \\
      \rowcolor{gray!20}
      56 & \textbf{0.70} & 19 & \textbf{0.76} & 17 & \textbf{0.67} &
      17 & \textbf{0.75} & 16 & \textbf{0.76} & 19 & \textbf{0.96} &
      17 & \textbf{0.93} & 16 & \textbf{0.95} & 20 \\

      \bottomrule
    \end{tabular}
  }
  \caption{Prediction performance across different targets as a
    function of selection window length for Comcast probes. Target locations are
    abbreviated by their first three letters. For each location,
    values in the \textbf{R} column
    denote recall while values in the \textbf{C} column denote
    probe counts. We observe invariance in recall scores across
    different selection window lengths. Additionally, for a given
    selection window, the recall scores follow a similar
    trend as the fraction of shared log-impact in
  Figure~\ref{fig:impact-sharing}.}
  \label{tab:training-period}
\end{table}

\paragraph{Experiment design.}
To assess whether our approach selects probes that continue to cover a
large fraction of anomalies over time, we perform a predictive
evaluation using a time-based split of the dataset. Specifically, we
use the first $T$ days to select probes via our greedy heuristic, and
evaluate the fraction of anomalies covered by these probes in the
remaining days. We vary $T$ from 7 to 56 days in increments of 7 days.
For each $T$, probe selection is re-optimized using only data from the
first $T$ days. We conduct this evaluation for multiple targets within the largest ISP
in our dataset (Comcast). To account for probe churn, we restrict
our focus to probes active in both the training and evaluation
periods, resulting in 29 eligible probes. For each $T$, we apply our
greedy heuristic to this set and report anomaly coverage in the
subsequent period.

\paragraph{Recall scores follow a similar trend as the fraction of shared
log-impact.} Table \ref{tab:training-period} shows the recall scores, i.e.,
fraction of anomalies covered in the future period by probes selected using data
from the selection window for different values of $T$. We exclude
precision because
it does not provide useful information since
there is no notion of false positives in this setting. A similar
argument applies for accuracy or F-1 score as they also rely on false
positives. We observe that the recall scores follow a similar
trend as the fraction of shared log-impact in Figure~\ref{fig:impact-sharing}.
For example, the recall scores peak at Stockholm, and finally drop for the
farthest targets (Tunis and Johannesburg). This suggests that the fraction of
shared log-impact between probes and targets is a good predictor of how well our
approach can select probes that cover future anomalies. In other
words, the ability
of a small-scale probe deployment to generalize to future anomalies
depends on the
location of the target being measured, with targets that are neither too
close nor too far away providing the best generalization performance.

\paragraph{Invariance to selection window length.}
Table~\ref{tab:training-period} also tells us that the length of the selection
window $T$ does not have a significant impact on the recall scores. This
suggests that even a small amount of historical data can be sufficient to select
probes that continue to provide a steady level of coverage of anomalies in the
future. This is a promising finding, potentially enabling operators,
researchers, or regulators who start with a small set of probes to experiment
with multiple deployment strategies, collect data for a short period of time,
and then use our approach to select a smaller subset of probes for long-term
deployment. An alternative approach could also be to start with a few probes and
incrementally add more to the deployment until a desired recall score is
achieved. We leave the exploration of such approaches to future work.


\section{Conclusion \& Future Directions} \label{sec:conclusion}

In this work, we apply change point detection to longitudinal idle latency measurements to characterize network anomalies by their amplitude and duration, and propose a topology-free probe selection algorithm that reduces redundant detections. In the absence of routing or path information, our approach exploits temporal overlap and similarity in anomaly magnitude across probes to identify a representative subset that preserves coverage of high-impact events. Our results show that shared anomalies provide a strong signal for distinguishing probes that contribute new information from those that observe largely redundant behavior.

At the same time, practical deployments exhibit constraints such as sparse probe density, incomplete longitudinal coverage, and lack of information about where additional measurements are most valuable. Our method relies on accurate anomaly detection methods to identify shared events across probes. While overlap thresholds and log-scaled impact scores mitigate change point detection noise, false positives can still affect probe selection. Moreover, the effectiveness of the approach depends on probe density, geographic and ISP diversity, and measurement frequency, which may vary across environments.

Based on these findings, we recommend that measurement platforms explicitly detect, share and leverage information shared anomalies when selecting probes. Future work should validate these techniques across multiple deployments, improve anomaly detection precision, and develop sampling strategies that better account for additional operational constraints. \label{LastPage}

\bibliographystyle{ACM-Reference-Format}
\bibliography{refs}

\clearpage

\appendix

\section{Ethics}

Our deployment of Raspberry Pi probes in residential networks was approved by our Institutional Review Board (IRB). During the deployment, we took extensive measures to preserve user privacy and ensure ethical data collection. Each installation of the Raspberry Pis was accompanied by a consent form that clearly outlined the purpose of data collection, the type of data being collected, and how the data would be used. No passive data was collected, and all data was anonymized before it was consumable for analysis. 

\section{Reasoning About Geographic Distribution of Probes} \label{sec:geo-dist}

\paragraph{Zip codes as an initial unit choice.} For a network
stakeholder designing a probe deployment, one important question may be the
following: \textit{with what geographic granularity should probes be
placed to achieve a desired coverage of anomalies?} To
attempt to answer this question, we analyze the geographic
distribution of probes selected by our approach relative to the
original probe distribution. We begin by comparing the overall probe
counts by zip codes in the original and reduced probe sets. We
initially pick zip codes as our default geographic granularity
because these are the immediate smaller geographic granularity that
appears in IP geolocation databases (e.g., MaxMind
\cite{maxmind_geolocation_database}, IPinfo \cite{ipinfo}) after
city-level geolocation. The zip code granularity further makes our
findings more applicable to a wide
range of measurement campaigns that may not have access to more
fine-grained geolocation data.

\begin{figure*}[t!]
  \centering
  \hfill
  \begin{subfigure}[t]{0.24\linewidth}
    \centering
    \captionsetup{width=.97\linewidth}
    \includegraphics[width=1.0\linewidth]{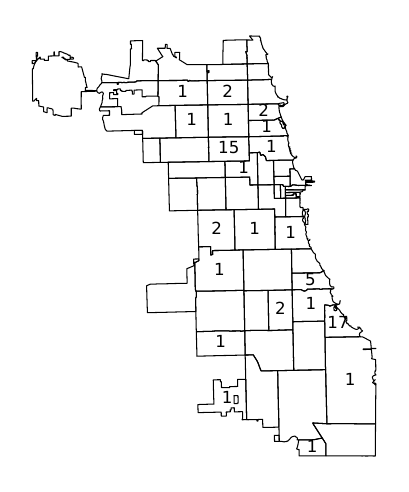}
    \caption{Original probe counts.}
    \label{fig:zipcode-distribution-original}
  \end{subfigure}
  \hfill
  \begin{subfigure}[t]{0.24\linewidth}
    \centering
    \captionsetup{width=.99\linewidth}
    \includegraphics[width=1.0\linewidth]{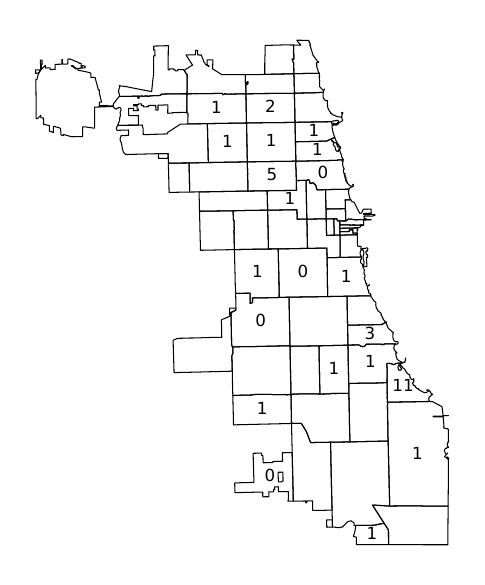}
    \caption{Reduced probe counts.}
    \label{fig:zipcode-distribution-reduced}
  \end{subfigure}
  \hfill
  \begin{subfigure}[t]{0.24\linewidth}
    \centering
    \captionsetup{width=.9\linewidth}
    \includegraphics[width=1.0\linewidth]{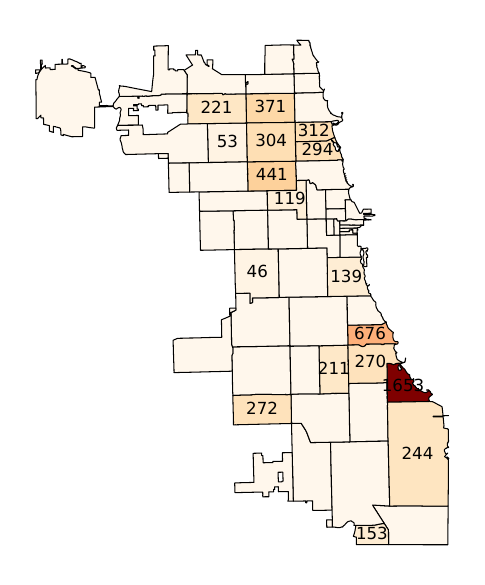}
    \caption{Number of unique anomalies covered by the reduced set.}
    \label{fig:spike-distribution-reduced}
  \end{subfigure}
  \hfill
  \begin{subfigure}[t]{0.24\linewidth}
    \centering
    \captionsetup{width=.9\linewidth}
    \includegraphics[width=1.0\linewidth]{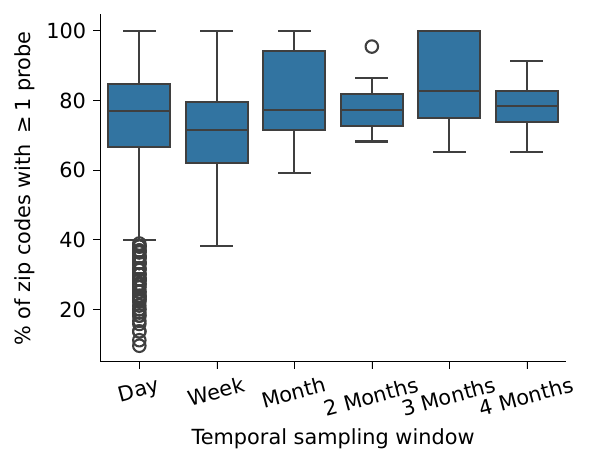}
    \caption{Impact of temporal sampling window.}
    \label{fig:zipcode-coverage-boxplot}
  \end{subfigure}
  \caption{Comparison of Comcast probe distribution by zip code in
    the original and reduced probe sets for the city of Chicago. 17
    of 21 zip codes retain
    at least one probe in the reduced set
    (\ref{fig:zipcode-distribution-original} and
    \ref{fig:zipcode-distribution-reduced}). A southern zip code
    contributes the most number of probes and unique anomalies
    (\ref{fig:spike-distribution-reduced}) from the reduced set. A
    high fraction of zip
    codes retain at least one probe across temporal splits of the
  dataset (\ref{fig:zipcode-coverage-boxplot}).}
  \label{fig:zipcode-distribution}
\end{figure*}

\paragraph{Implicit geographic diversity.} Figure~\ref{fig:zipcode-distribution}
shows a comparison of the original and reduced probe distributions by zip code
for the city of Chicago\footnote{Two probes are intentionally omitted due to
being located outside  city boundaries.}. We calculate the reduced set by
continuing with our choice of $\delta_{IoU}$ as 0.9, and using all probes that
were active throughout our deployment. We then calculate the total number of
probes required to cover 95\% of anomalies. The reduced set results in a total
of 33 probes, one of which is located in a previously unmonitored zip code. A
total of 21 zip codes are shown in
Figure~\ref{fig:zipcode-distribution-original}, out of which 17 zip codes
(80.9\% of total) retain at least one probe in the reduced set
(Figure~\ref{fig:zipcode-distribution-reduced}). Our greedy heuristic appears to
implicitly preserve the geographic diversity of the probes from the original
set, as suggested by the majority of zip codes retaining at least one probe in
the reduced set. This further indicates that a geographically diverse probe
deployment is important to achieve a high coverage of anomalies for one ISP,
\textit{even} when all probes are located within the same city. Finally, we also
observe that a southern zip code (60649) contributes the most number of probes
(11) and unique anomalies (1653) from the reduced set. This zip code overlaps
with the South Shore neighborhood which is known to be racially segregated
\cite{sharma2022benchmarks}, suggesting that probes deployed in
socioeconomically disadvantaged areas may experience greater heterogeneity in
network performance that is not captured by probes elsewhere.

\paragraph{Validation using resampling.} To validate that this
observation is not an artifact of the underlying data, we also run
our greedy heuristic on time-based splits of our dataset using the
same parameters as above ($\delta_{IoU} = 0.9$, $c = 1.0$, towards
Chicago). To this end, we
resample our dataset into
daily, weekly, monthly, bi-monthly and quarterly bins and run our
heuristic on sliding windows of these bins. For each window, we
calculate the fraction of zip codes that preserve at least one probe
in the reduced set. Figure~\ref{fig:zipcode-coverage-boxplot} shows a
box plot of this fraction versus the sampling window size. We observe
a high variation in this metric across daily and weekly windows,
which is expected due to smaller sample sizes and churn in the probe
population. However, for sampling window sizes of a month or more we
observe a median fraction of $>0.77$ with low variation, suggesting
significant robustness of our observation to the underlying data.

\paragraph{Comparison with other geographic granularities.} We also
look at additional geographic granularity choices including Chicago's
census tracts and neighborhood boundaries. For the complete dataset,
we estimate the \textit{fraction of geographic regions that retain at
least one probe} in the reduced set for each choice of geographic
granularity. We estimate this metric to be 77.64\% $\pm$ 7.44\% for
zip codes, 4.76\% $\pm$ 0.00\% for neighborhoods, and only 2.32\%
$\pm$ 0.14\% for census tracts, where the $\pm$ values represent one
standard deviation. Zip codes, being the larger geographic regions,
have a high fraction of regions retaining at least one probe. The low
fractions for neighborhoods and census tracts suggest that
measurement sampling designers should first consider placing multiple
probes at a granularity as large as zip codes before expanding to
finer granularities. From Figure~\ref{fig:zipcode-distribution}, we
also observe that the two zip codes with 15 and 17 probes in the
original set retain 5 and 11 probes respectively in the reduced set.
This suggests that some zip codes may require more probes than others
to achieve a satisfactory level of anomaly coverage. Prior to actual
probe deployments, crowdsourced datasets such as M-Lab
\cite{mlab2025} or Ookla \cite{ookla2025} could be used to identify
such areas based on the testing density of users as well as the
variability in network performance experienced by these users. Given these
insights, we believe that a future research direction could be to
combine these data
sources with our approach for arriving at more informed probe
placement strategies.

\end{document}